\documentclass{article}
\usepackage{spconf,amsmath,graphicx,amsfonts,float,commath}
\usepackage{xcolor}

\title{Efficient Trainable Front-Ends for Neural Speech Enhancement}
\name{Jonah Casebeer\sthanks{Work performed while at Amazon Web Services.}$^{\dagger}$ \qquad Umut Isik$^{\sharp}$ \qquad Shrikant Venkataramani\footnotemark[1]$^{\dagger}$ \qquad Arvindh Krishnaswamy$^{\sharp}$}

\address{$^{\sharp}$ Amazon Web Services, Amazon AI  \\
         $^{\dagger}$ University of Illinois at Urbana-Champaign\\
        }

\begin{document}
\ninept
\maketitle
\begin{abstract}
Many neural speech enhancement and source separation systems operate in the time-frequency domain. Such models often benefit from making their Short-Time Fourier Transform (STFT) front-ends trainable. In current literature, these are implemented as large Discrete Fourier Transform matrices; which are prohibitively inefficient for low-compute systems. We present an efficient, trainable front-end based on the butterfly mechanism to compute the Fast Fourier Transform, and show its accuracy and efficiency benefits for low-compute neural speech enhancement models. We also explore the effects of making the STFT window trainable.  

\end{abstract}
\section{Introduction}
The performance of speech enhancement and source separation systems has vastly improved with the introduction of deep learning and neural network based techniques~\cite{huang2014deep, venkataramani2018end, luo2018tasnet, luo2019conv, shi2019deep, rethage2018wavenet, hershey2016deep, yu2017permutation, weninger2014discriminatively}. Some recent advances include the use of generative adversarial networks~\cite{pascual2017segan}, sophisticated adaptations of U-Net \cite{ronneberger2015u} based architectures \cite{ macartney2018improved, stoller2018wave} and many more. Designing end-to-end systems that directly estimate the waveforms of the enhanced speech signal by operating on the noisy speech waveforms has proven to be beneficial and resulted in several high-performance models \cite{venkataramani2018end, luo2018tasnet, luo2019conv, shi2019deep, stoller2018wave, rethage2018wavenet}.

End-to-end speech enhancement networks typically replace the Short-Time Fourier transform (STFT) operation by a learnable 'front-end' layer \cite{venkataramani2018end, luo2018tasnet, luo2019conv, shi2019deep}. To this end, the first layer of such neural models performs windowing followed by a dense matrix multiplication. To transform the data back into the waveform domain, these models also employ a trainable back-end layer which inverts the front-end via another dense matrix multiplication and the overlap-add method.


With growing interest in ``hearables'' and other wearable audio devices, low-compute and real-time scenarios are increasingly encountered in audio processing applications. These devices come with low power, low memory and stringent compute requirements but offer the opportunity for audio processing everywhere. In these contexts, storing and performing inference with dense matrices inside a trainable STFT can be prohibitively expensive or downright infeasible. For example, to learn an STFT with $512$-point transforms takes $2 \times 512 \times 512 = 524288$ parameters. The front-end parameters alone could fill the L2 cache of a modern processor, leaving no room for the rest of the model to be evaluated without cache-misses. We aim to address this issue by creating an efficient front-end for low-compute models operating directly on the waveform.

In this work we propose an efficient learnable STFT front-end for low-compute audio applications and show how it can be used to improve end-to-end speech enhancement. The trainable FFT copies the butterfly mechanism for the Fast Fourier Transform (FFT); and, when initialized appropriately, computes the Fourier Transform. We also propose replacing the standard fixed window of the STFT by a trainable windowing layer. In terms of computational advantages, our model requires no increase in compute and a minimal increase in the number of required parameters compared to the fixed FFT. Using our model also leads to significant savings in memory compared to standard adaptive front-end implementations. We evaluate our model on the VCTK speech enhancement task \cite{valentini2017noisy} and demonstrate that the proposed front-end outperforms STFT front-ends on a variety of perceptually-motivated speech enhancement metrics. 

\section{The FFT as a Network Layer}
Fast Fourier Transforms (FFT) are based on factoring the Discrete Fourier Transform (DFT) matrix into a set of sparse matrices \cite{van1992computational}. We implement these sparse matrix operations efficiently in MXNet \cite{chen2015mxnet} to make a trainable layer for low-compute environments. The factorization we use is based on the butterfly mechanism and is as follows.


Recall that, given a vector $\mathbf{x}$ of length $N$, the $N$-point DFT applies a transformation $\mathbf{F}_{N}$ to get the DFT  coefficients $\mathbf{X}$; with the element-wise version of this operation being,

\begin{equation}
    \mathbf{X}[k] = \sum \limits_{n=0}^{N-1} \mathbf{x}[n] \cdot \omega_{N}^{kn}
    \label{eq:dft}
\end{equation}

Here, $\omega_N = e^{\frac{-2 \pi j}{N}}$ denotes the twiddle-factor of the DFT. As usual, split Eq.~\ref{eq:dft} into two $\frac{N}{2}$ point DFT operations on the even-indexed and the odd-indexed elements of $\mathbf{x}$,

\begin{align}
    \mathbf{X}[k] &= \sum \limits_{n=0}^{\frac{N}{2}-1} \mathbf{x}[2n] \omega_{\frac{N}{2}}^{kn} + \omega_N^k \sum \limits_{n=0}^{\frac{N}{2}-1} \mathbf{x}[2n+1] \omega_{\frac{N}{2}}^{kn} \\
    &= \mathbf{E}[k] + \omega_N^k \cdot \mathbf{O}[k]
    \label{eq:dft_even_odd_split}
\end{align}

The twiddle-factors are odd-symmetric about $k=\frac{N}{2}$ i.e., $\omega_{N}^{\left(k+\frac{N}{2}\right)} = - \omega_{N}^{k}$. Thus, 

\begin{equation}
    \mathbf{X}[k] = 
    \begin{cases}
        \mathbf{E}[k] + \omega_N^k \mathbf{O}[k] & 0 \leq k < \frac{N}{2} \\
        \mathbf{E}[k-\frac{N}{2}] - \omega_N^{k-\frac{N}{2}} \mathbf{O}[k-\frac{N}{2}] & \frac{N}{2} \leq k < N 
    \end{cases}
    \label{eq:dft_twiddle_symmetric}
\end{equation}

Defining a diagonal matrix of twiddle-factor values, $\mathbf{\Omega}_\frac{N}{2} = \operatorname{diag}\{\omega_N^0,\omega_N^1,\cdots,\omega_N^{\frac{N}{2}-1}\}$, we rewrite Eq.~\ref{eq:dft_twiddle_symmetric} in matrix form as,

\begin{equation}
    \mathbf{X} = 
    \begin{bmatrix}
        \mathbf{F}_\frac{N}{2} \cdot \mathbf{x}_{\text{even}} + \mathbf{\Omega}_\frac{N}{2} \cdot \mathbf{F}_\frac{N}{2} \cdot \mathbf{x}_{\text{odd}}\\
        \mathbf{F}_\frac{N}{2} \cdot \mathbf{x}_{\text{even}} -\mathbf{\Omega}_\frac{N}{2} \cdot \mathbf{F}_\frac{N}{2} \cdot \mathbf{x}_{\text{odd}}
    \end{bmatrix}
\end{equation}
In this equation, $\mathbf{x}_{\text{odd}}$ and $\mathbf{x}_{\text{even}}$ denote the odd-indexed and even-indexed terms of $\mathbf{x}$.

\begin{equation}
    \mathbf{X} = 
    \begin{bmatrix}
        \mathbf{F}_\frac{N}{2}  & \mathbf{\Omega}_\frac{N}{2} \cdot \mathbf{F}_\frac{N}{2}\\
        \mathbf{F}_\frac{N}{2}  & -\mathbf{\Omega}_\frac{N}{2} \cdot \mathbf{F}_\frac{N}{2}
    \end{bmatrix}
    \cdot
    \begin{bmatrix}
         \mathbf{x}_{\text{even}}\\
         \mathbf{x}_{\text{odd}}
    \end{bmatrix}
\end{equation}

Then, we factor out the $\frac{N}{2}$-point DFT, and apply an even/odd permutation matrix $\mathbf{P}_N$ to get,
\begin{equation}
    \mathbf{X} = 
    \begin{bmatrix}
        \mathbf{I}_\frac{N}{2}  & \mathbf{\Omega}_\frac{N}{2}\\
        \mathbf{I}_\frac{N}{2}  & -\mathbf{\Omega}_\frac{N}{2}
    \end{bmatrix}
    \begin{bmatrix}
        \mathbf{F}_\frac{N}{2}  & 0\\
        0  & \mathbf{F}_\frac{N}{2}
    \end{bmatrix}
    \cdot \mathbf{P}_{N} \cdot \mathbf{x}
\end{equation}

Disregarding the data, we can write $\mathbf{F}_N$ as,

\begin{equation}
    \mathbf{F}_N = 
    \begin{bmatrix}
        \mathbf{I}_\frac{N}{2}  & \mathbf{\Omega}_\frac{N}{2}\\
        \mathbf{I}_\frac{N}{2}  & -\mathbf{\Omega}_\frac{N}{2}
    \end{bmatrix}
    \begin{bmatrix}
        \mathbf{F}_\frac{N}{2}  & 0\\
        0  & \mathbf{F}_\frac{N}{2}
    \end{bmatrix}
    \cdot \mathbf{P}_{N}
\end{equation}.

Substitute $\mathbf{W}_m$ for the matrix of twiddle factors to get,

\begin{equation}
    \mathbf{F}_N = \mathbf{W}_m \cdot
        \begin{bmatrix}
            \mathbf{F}_\frac{N}{2}  & 0\\
            0  & \mathbf{F}_\frac{N}{2}
        \end{bmatrix}
    \cdot \mathbf{P}_{N}
\end{equation}

To simplify even further, $\mathbf{F}_{\frac{N}{2}}$ is,

\begin{equation}
    \mathbf{F}_\frac{N}{2} =
    \cdot
    \begin{bmatrix}
        \mathbf{I}_\frac{N}{4}  & \mathbf{\Omega}_\frac{N}{4}\\
        \mathbf{I}_\frac{N}{4}  & -\mathbf{\Omega}_\frac{N}{4}
    \end{bmatrix}
    \begin{bmatrix}
        \mathbf{F}_\frac{N}{4}  & 0\\
        0  & \mathbf{F}_\frac{N}{4}
    \end{bmatrix}
    \cdot \mathbf{P}_{\frac{N}{2}}
\end{equation}

Thus, we can write $\mathbf{F}_N$ in terms of $\mathbf{F}_{\frac{N}{4}}$ as,

\begin{multline}
    \mathbf{F}_N = \mathbf{W}_m \cdot \mathbf{W}_{m-1}
    \cdot
    \begingroup
    \setlength\arraycolsep{.1pt}
    \begin{bmatrix}
        \mathbf{F}_\frac{N}{4} &&&\\
        & \mathbf{F}_\frac{N}{4}&&\\
        && \mathbf{F}_\frac{N}{4}&\\
        &&& \mathbf{F}_\frac{N}{4}
    \end{bmatrix}
    \cdot
    \begin{bmatrix}
        \mathbf{P}_\frac{N}{2} &\\
        & \mathbf{P}_\frac{N}{2}\\
    \end{bmatrix}
    \endgroup
    \cdot
    \mathbf{P}_N
\end{multline}

where,
\begin{equation}
\mathbf{W}_{m-1} = 
    \begin{bmatrix}
        \mathbf{I}_\frac{N}{4} & \mathbf{\Omega}_\frac{N}{4}&&\\
        \mathbf{I}_\frac{N}{4} & -\mathbf{\Omega}_\frac{N}{4}&&\\
        && \mathbf{I}_\frac{N}{4} & \mathbf{\Omega}_\frac{N}{4} \\
        && \mathbf{I}_\frac{N}{4} & -\mathbf{\Omega}_\frac{N}{4}
    \end{bmatrix}
\end{equation}

\begin{figure}[ht]
    \centering
    \includegraphics[scale=0.8]{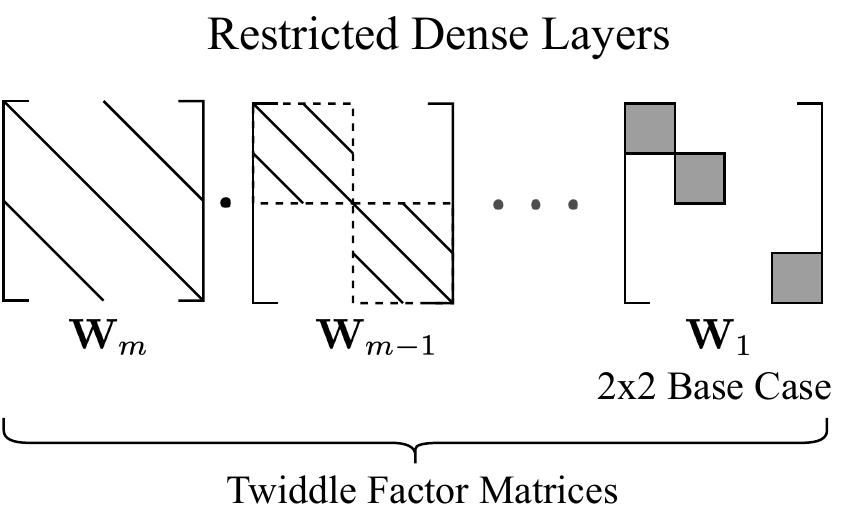}
    \vspace{-3mm}
    \caption{The DIT-FFT algorithm computes the Fourier transform as a series of sparse matrix multiplies. This figure demonstrates the structure involved in these matrices. The solid diagonal lines indicate the positions of the non-zero elements in the matrices, and the grey squares represent $2 \times 2$ DFT matrices.}
    \label{fig:fft_sketch}
\end{figure}

It is necessary to stack the result of $\mathbf{F}_{\frac{N}{2}}$ since it occurs more than once. The component matrices $\mathbf{W}_m$ and $\mathbf{W}_{m-1}$ are composed of stacks of diagonal matrices. Generalizing this further, we can represent an $N$-point FFT as a series of $\log_2(N)+1$ matrix multiplications where, the first matrix is a permutation matrix and all other matrices are sparse matrices formed by stacks of diagonal matrices. Mathematically, we can write the $N$-point DIT-FFT algorithm as a matrix multiplication by 
\begin{equation}
    \mathbf{F}_{N} = \mathbf{W}_{m} \cdot \mathbf{W}_{m-1} \cdots \mathbf{W}_{1} \cdot \mathbf{B}_{N}
\end{equation}
where, $\mathbf{W}_{m}$ denotes the $m$'th twiddle factor matrix, $\mathbf{B}_{N}$ denotes the product of all permutation matrices $\mathbf{P}$, and $m = \log_2(N)$ is the number of twiddle factor matrix multiplications involved. We can write a general formula to construct the $k^{\text{th}}$ twiddle factor matrix $\mathbf{W}_k$ of the $m$ twiddle factor matrices using identity matrices $\mathbf{I}_{\frac{N}{2^k}}, \mathbf{I}_{2^{k-1}}$ of size $\frac{N}{2^k}$ and $2^{k-1}$, as well as the Kronecker product $\otimes$ as follows, 

\begin{equation}
    \mathbf{W}_{k} =\mathbf{I}_{\frac{N}{2^k}} \otimes \left[
    \begin{array}{cc}
        \mathbf{I}_{2^{k-1}} &  \mathbf{\Omega}_{2^{k-1}} \\ 
        \mathbf{I}_{2^{k-1}} & -\mathbf{\Omega}_{2^{k-1}}
    \end{array} \right]\\
\end{equation}
Fig. \ref{fig:fft_sketch} visualizes these matrices and the associated sparsity patterns for an $N$-point DIT-FFT algorithm..

As an illustrative example, consider the FFT matrix derivation for a $4$-point FFT operation. Given the permuted data samples, we first apply a $2$-point DFT operation on successive pairs of input samples. This step can be written as a matrix multiplication operation by matrix $\mathbf{W}_{1}$ where, 

\[\textit{$\mathbf{W}_{1}$} =
\begin{bmatrix}
    1  & 1    & 0    & 0 \\
    1  & -1    & 0    & 0 \\
    0  & 0    & 1    & 1 \\
    0  & 0    & 1    & -1
\end{bmatrix}\]

The next step applies a matrix multiplication operation using matrix $\mathbf{W}_{2}$ where we can write $\mathbf{W}_{2}$ in terms of the twiddle factor $\omega_{4} = e^{\frac{-2 \pi j}{4}}$ as,  
\[\textit{$\mathbf{W}_{2}$} =
\begin{bmatrix}
    1  & 0    & \omega_4^0    & 0 \\
    0  & 1    & 0    & \omega_4^1 \\
    1  & 0    & -\omega_4^0    & 0 \\
    0  & 1    & 0    & -\omega_4^1
\end{bmatrix}\]

The overall 4-point FFT operation can be expressed as,
\begin{equation*}
    \mathbf{X} = \mathbf{W}_{2} \cdot \mathbf{W}_{1} \cdot \mathbf{B}_{4} \cdot \mathbf{x}
\end{equation*}
We see that the matrix $\mathbf{W}_{1}$ is a sparse matrix with a block diagonal structure. Similarly, $\mathbf{W}_{2}$ is also sparse and composed of stacks of diagonal matrices. 

\subsection{Trainable FFT layer}
\begin{figure}[ht]
    \centering
    \includegraphics[scale=0.7]{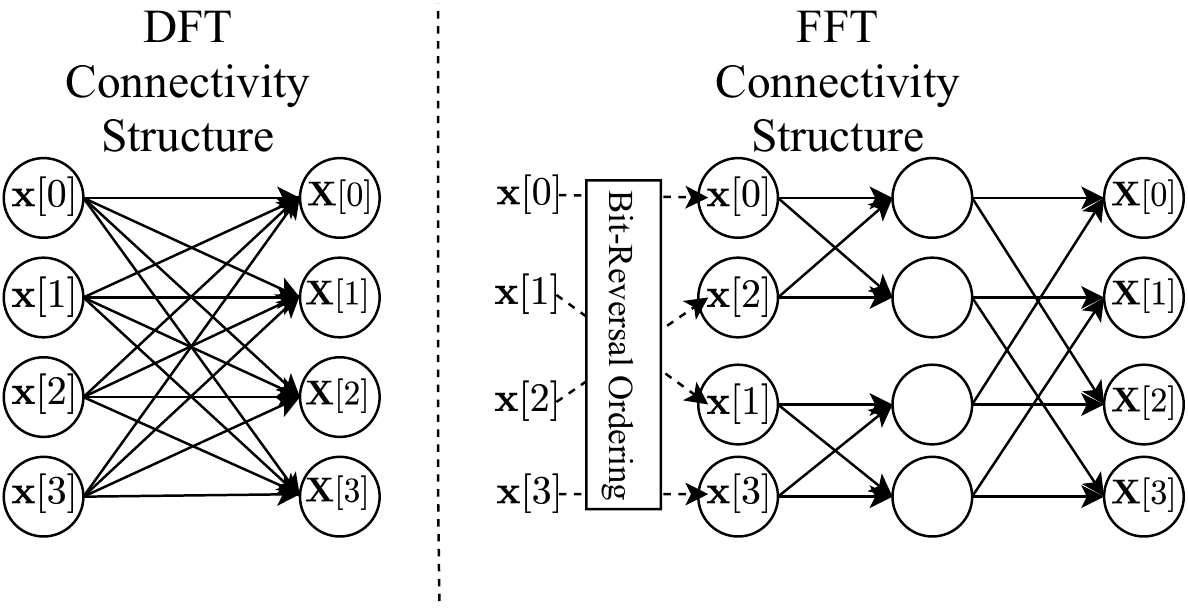}
    \vspace{-4mm}
    \caption{On the left we show the connectivity pattern enforced by a DFT matrix. It is a dense linear layer since the DFT matrix application can be represented by a single dense matrix multiply. On the right we show the sparser connectivity pattern and bit-reversal enforced by the DIT-FFT algorithm. The FFT connectivity structure makes all matrices in the FFT factorization trainable except for the bit-reversal permutation matrix.}
    \label{fig:butterfly}
    \vspace{-4mm}
\end{figure}

To make the above FFT layer trainable, the set of matrix multiplies can be represented as a neural network with several sparsely connected layers. The DFT on the other hand is a single layer with dense connectivity. We preserve the general FFT structure by only connecting network nodes on the block diagonal structure given by the FFT. This preserves the speed and structure of the FFT while both reducing the number of parameters in the front-end and operating on the raw waveform. Fig.~\ref{fig:butterfly} illustrates the DFT and FFT connectivity structures and how they may be interpreted as neural network layers. In practice, we explicitly implement all complex operations with real values. When initialized to do so, the model returns identical results to the FFT algorithm. All of these operations can be efficiently implemented and trained with sparse matrix routines.


\subsection{Inverse FFT}
To compute the inverse FFT we use the time reversal conjugate trick. Given the DFT representation $\mathbf{X}$ of the time domain frame $\mathbf{x}$, we can compute the inverse Fourier transform of $\mathbf{X}$ using only the FFT. In particular, 
\[\mathbf{x}=\frac{\overline{\operatorname{FFT}(\overline{\mathbf{X}}))}}{N}\]

In our model we leverage this property. However, the forward FFT layer and the inverse FFT layer do not share parameters. We use different learned FFTs for the forward and inverse transforms. These FFT layers are initialized and updated as separate entities.

\section{The Adaptive STFT for Speech Enhancement}
Given a trainable FFT front-end, we can now operate the model on time domain waveforms. For our models we use 256-point FFTs such that this model's front-end has about two orders of magnitude fewer weights than a typical trainable STFT front-end. In addition to making the FFT and IFFT trainable, we also show how we can make trainable synthesis and analysis windows. 


\subsection{Learning a Window}
The typical $N$-point STFT with a hop-size of $h$, chunks an input audio sequence $\mathbf{x}$ into windows of size $N$ with an overlap of $N-h$. These overlapping frames are stacked to form columns of a matrix. Call this matrix $\mathbf{S}$ where the first column of $\mathbf{S}$ holds values $\mathbf{x}[0:N-1]$, the second column holds values $\mathbf{x}[h:h+N-1]$, and so on in the standard python notation. To apply a windowing function upon $\mathbf{S}$, construct a windowing matrix $\mathbf{H}$. The matrix $\mathbf{H}$ is diagonal with the desired $N$-point window on its diagonal. The windowed version of $\mathbf{S}$ is then $\mathbf{S}_\text{win}=\mathbf{H} \cdot \mathbf{S}$. This same logic applies for windowing in the inverse STFT. 

\begin{figure}[ht]
    \centering
    \includegraphics[scale=0.4]{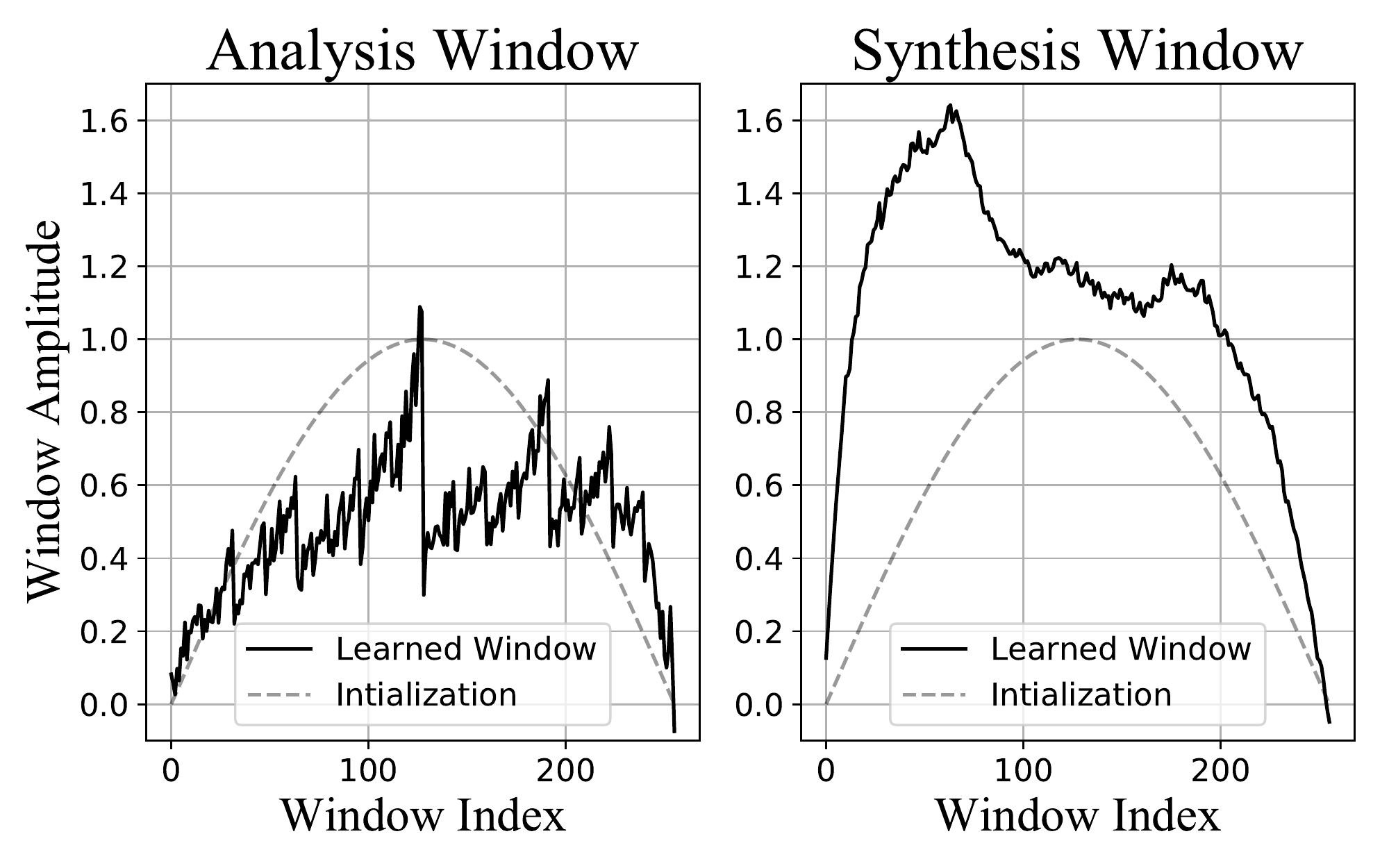}
    \vspace{-4mm}
    \caption{The left plot contains the analysis window. It has fairly regular high frequency patterns. The right plot shows the synthesis window. Interestingly, it is two peaked. Both windows have changed considerably from their intializations.}
    \label{fig:windows}
    \vspace{-4mm}
\end{figure}

By making $\mathbf{H}$ a network parameter, we can learn a windowing function suited to the learned transform. For all our experiments, $\mathbf{H}$ is initialized as a Hann window. During training, $\mathbf{H}$ is freely updated without non-negativity or constant-overlap-add constraints. In Fig.~\ref{fig:windows}, we show that the learned analysis and synthesis windows are highly structured.

\subsection{Model Architecture}
In conjunction with the learned transforms and windows, we use a masking based separation network. The learned FFT front-end transforms each column of $\mathbf{S}_\text{win}$. Let $\operatorname{FFT}_{train}(\cdot)$ represent the trainable FFT front-end. The masking network takes $\operatorname{FFT}_{train}(\mathbf{S}_\text{win})$ and predicts two sigmoid masks: $\mathbf{M}_r$ and $\mathbf{M}_i$. These masks are applied via element-wise multiplication to produce an estimate of the clean speech $\widehat{\mathbf{S}}_{clean}$ in the transform domain, in term of its real and imaginary parts. Specifically,

\begin{equation}
    \begin{aligned}
        \operatorname{Re}(\widehat{\mathbf{S}}_{clean}) = \operatorname{Re}(\operatorname{FFT}_{train}(\mathbf{S}_\text{win})) \odot \mathbf{M}_r\\
        \operatorname{Im}(\widehat{\mathbf{S}}_{clean}) = \operatorname{Im}(\operatorname{FFT}_{train}(\mathbf{S}_\text{win})) \odot \mathbf{M}_i
    \end{aligned}
\end{equation}

\begin{figure}[h]
    \centering
    \includegraphics[scale=0.84]{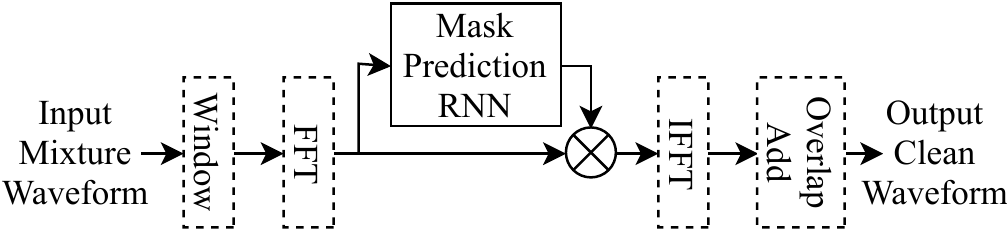}
    \vspace{-4mm}
    \caption{This block diagram shows the full pipeline for the proposed model. Inside the solid black box are the operations which are trained when using a fixed transform. The dashed boxes contain the additional operations trained in our setup.}
    \label{fig:model_pipeline}
\end{figure}

Here $\operatorname{Re}(.)$ and $\operatorname{Im}(.)$ compute the element-wise real and imaginary components of a complex matrix, and $\odot$ denotes the element-wise multiplication operation. In our experiments, we found that using separate real and imaginary masks outperformed a single magnitude mask. Fig.~\ref{fig:model_pipeline} illustrates the full model pipeline.

\begin{figure}[h]
    \centering
    \includegraphics[scale=.8]{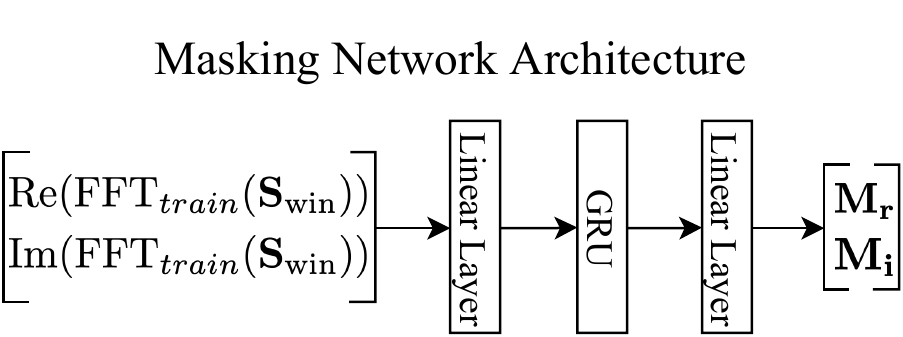}
    \vspace{-4mm}
    \caption{The masking network is composed of three layers: two linear layers, and a unidirectional gated recurrent unit layer. This network is causal for real-time speech enhancement. The real and imaginary components of the transformed input are stacked before being fed through the network. Similarly, the output is interpreted as the real mask stacked on top of the imaginary mask.}
    \label{fig:mask_rnn}
    \vspace{-5mm}
\end{figure}

We experimented with a mask prediction RNN containing 80k parameters. This network is composed of two linear layers, and a gated recurrent unit (GRU). The GRU is unidirectional for our intended use case of real-time speech enhancement. Instead of performing complex valued back-propagation, we simply stack the real and imaginary components of the input and output. The masking network architecture is shown in Fig.~\ref{fig:mask_rnn}.

\section{Experiments}

\subsection{Dataset}
We use the 56 speaker VCTK training and testing setup \cite{valentini2017noisy} where, each speaker has about 400 sentences. During training, we mix speech and noise at signal to noise ratios~(SNRs) of 0dB, 5dB, 10dB, and 15dB. We train all models until convergence on the training set before evaluating them. During evaluation, we use SNRs of 2.5dB, 7.5dB, 12.5dB, and 17.5dB  \cite{valentini2016investigating}.


\subsection{Training}

We used the MXNet \cite{chen2015mxnet} framework for all our experiments. To optimize the network parameters, we use the Adam algorithm \cite{kingma2014adam}, and for the loss function, we use the complex loss $\mathcal{L}$ given in Eq.~\ref{eq:loss_function}~\cite{wilson2018exploring}. Over informal experiments, we found that this loss function performed better than time domain loss functions in terms of perceptual metrics. The loss function is a weighted combination of magnitude mean squared error and complex mean squared error loss. Here, $\widehat{\mathbf{Y}}$ is the predicted clean Fourier spectrum and $\mathbf{Y}$ is the true clean Fourier spectrum.

\begin{equation}
    \mathcal{L} = \| |\widehat{\mathbf{Y}}|^{\alpha}  - |\mathbf{Y}|^{\alpha} \|_2^2 + \lambda \| \widehat{\mathbf{Y}}^{\alpha}  - \mathbf{Y}^{\alpha}  \|_{2}^{2}
\label{eq:loss_function}
\end{equation}

The power $\alpha$ is applied element-wise and in the case of complex numbers is applied on the predicted magnitude and then multiplied with the predicted phase. For our own experiments, we use $\alpha=0.3$ and $\lambda=0.1$ \cite{wilson2018exploring}.


\subsection{Models}
With low-compute scenarios in mind, we examined a model with approximately 80k parameters. The majority of the parameters are used in the masking network with the learned FFT and learned window using 512 parameters each. For all model runs, we initialize the windows as Hann windows and the trainable FFTs as FFTs. In our experiments, we compared four models with the attributes described below. The tested setups are: (1) Fixed Window Fixed FFT, (2) Trainable Window Fixed FFT, (3) Fixed Window Trainable FFT, (4) Trainable Window Trainable FFT. In the above list, fixed denotes parameters that were frozen and not updated during training. The Fixed Window Fixed FFT model has $\approx 1\%$~(1024) fewer parameters than the Trainable Window Trainable FFT model.



The first model has fixed Hann windows and a fixed FFT. This model only learns a masking network and serves as a benchmark for our adaptations. The other models serve to illustrate the benefits of and relationship between trainable windows and trainable FFTs.

\subsection{Evaluation Metrics}
We evaluate the above models on a speech enhancement task and compare them on the following metrics: signal distortion ($C_{sig}$), noise distortion ($C_{bak}$), overall quality ($C_{ovl}$) \cite{hu2006evaluation}, Perceptual Evaluation of Speech Quality (PESQ) \cite{recommendation2001perceptual}, and segmental SNR (SSNR). $C_{sig}$, $C_{bak}$, $C_{ovl}$, and PESQ are perceptual metrics intended to imitate a mean opinion score test. $C_{sig}$ estimates distortion of the speech signal, $C_{bak}$ estimates intrusiveness of the noise, $C_{ovl}$ summarizes the overall quality, and PESQ estimates the speech quality. For all of these metrics, higher is better.

\subsection{Results}
\begin{table}
    \begin{center}
            \begin{tabular}{||c |c c c c||} 
            \hline
            Trainable Window &$\times$ &\checkmark &$\times$      &\checkmark\\
            \hline
            Trainable FFT    &$\times$ &$\times$     &\checkmark  &\checkmark\\
            \hline\hline
            $C_{sig}$ &    3.586 &  3.580  &  3.624  &  \textbf{3.686}\\ 
            \hline
            $C_{bak}$ &    2.820 &  2.791  &  2.868  &   \textbf{2.942}\\
            \hline
            $C_{ovl}$ &    2.878 &  2.868  &  2.944  &   \textbf{3.018}\\
            \hline
            PESQ &    2.217 &  2.204  &  2.312  &   \textbf{2.395}\\
            \hline
            SSNR &    5.572 &  5.256  &  5.657  &   \textbf{6.137}\\
            \hline
            LOSS &    0.079 &  0.080  &  0.071  &  \textbf{0.070}\\
            \hline
            \end{tabular}
    \end{center}
    \vspace{-4mm}
    \caption{Comparison of speech enhancement performance on the VCTK test set using several perceptual metrics. We compare performance between several front-end setups. For the trainable window/FFT attributes we use \checkmark when this attribute was trainable and $\times$ when it was not. In this results table we include the loss as defined in the training section. For the loss, lower is better. Finally, the best score for each metric is displayed in bold.}
    \label{Tab:results}
\end{table}


Table~\ref{Tab:results} gives the results of our experiments for the 80k parameter model. We use the fixed FFT, fixed window version without any trainable parameters in its front-end as the baseline model. In general, we observe that making the FFT layer trainable improves speech enhancement performance. This improvement is consistently observed both in the case of a fixed window (compare column-$1$ to column-$3$) and when the window is trainable (compare column-$2$ to column-$4$). The effects of the window function are more inconsistent and interesting. A trainable window with a fixed FFT degrades separation performance (compare column-$1$ to column-$2$). Alternatively, a trainable window used with a trainable FFT improves upon the fixed window, trainable FFT model (compare column-$3$ to column-$4$). Overall, when a trainable window is used in conjunction with a trainable FFT, we get the best performance across all metrics.

\section{Conclusion}
In light of the need for high-performance speech systems in low-compute contexts, we proposed an alternative to learning a DFT matrix in trainable STFT systems. Our efficient front-end leverages the sparse structure of FFTs to both reduce computational and memory requirements. The trainable front-end is made up of several highly structured sparse linear layers and a learned window. We demonstrate an application of this front-end in speech enhancement. Using a trainable FFT and a trainable window improves speech enhancement performance over a fixed STFT system with no increase in computational complexity.

\bibliographystyle{IEEEbib}
\bibliography{strings,refs}

\end{document}